\documentclass[aps,pre,twocolumn,groupedaddress,showpacs]{revtex4-1}
\usepackage[english]{babel}
\usepackage[latin1]{inputenc}
\usepackage{setspace}
\usepackage{graphicx,times}
\usepackage{amsmath,amssymb}

\usepackage{pstricks,pst-grad}

\begin{document}
\title{Asymmetric coevolutionary voter dynamics}
\author{Stefan Wieland}
\email{swieland@cii.fc.ul.pt}
\author{Ana Nunes}
\email{anunes@ptmat.fc.ul.pt}
\affiliation{Centro de F{\'\i}sica da Mat{\'e}ria Condensada and Departamento de F{\'\i}sica, Faculdade de Ci{\^e}ncias da Universidade de Lisboa, P-1649-003 Lisboa, Portugal}

\begin{abstract}
We consider a modification of the adaptive contact process which, interpreted in the context of opinion dynamics, breaks the symmetry of the coevolutionary voter model by assigning to each node type a different strategy to promote consensus: orthodox opinion holders spread their opinion via social pressure and rewire their connections following a segregationist strategy; heterodox opinion holders adopt a proselytic strategy, converting their neighbors through personal interactions, and relax to the orthodox opinion according to its representation in the population. We give a full description of the phase diagram of this asymmetric model, using the standard pair approximation equations and assessing their performance by comparison with stochastic simulations. 
We find that although global consensus is favored with regard to the symmetric case, the asymmetric model also features an active phase. We study the stochastic properties of the corresponding metastable state in finite-size networks, discussing the applicability of the analytic approximations developed for the coevolutionary voter model. We find that, in contrast to the symmetric case, the final consensus state is predetermined by the system's parameters and independent of initial conditions for sufficiently large system sizes. We also find that rewiring always favors consensus, both by significantly reducing convergence times and by changing their scaling with system size.
\end{abstract}
\maketitle
\section{Introduction}
Many collective phenomena arise through structured interactions among a system's constituents \cite{Dorogovtsev2003,NewmanSIAM2003}. In this network of interactions, nodes (representing the system's agents) are assigned states and links to other nodes, with node states driving system dynamics along links. The coupling of node-state dynamics with their topological background is a widespread occurence and has been explored computationally as well as analytically in the field of adaptive networks \cite{Blasius2008,Gross2009,Sayama2013}.

The long-term behavior is an important characteristic of adaptive networks and usually assessed by low-dimensional systems of ordinary differential equations (ODEs) for the evolution of a number of network motifs' densities, obtained through mean-field approximations
 \cite{VazquezNJP2008,GrossPRL2006,KimuraPRE2008, BoehmePRE2011}. Possible asymptotic regimes either feature static state and link network configurations (\emph{frozen state}) or perpetual dynamics in the thermodynamic limit (\emph{active state}). Frozen states are further classified according to how dynamics come to a halt. At absorbing consensus, a single node state is adopted globally and updating stops because of node-state homogeneity. The absorbing fragmented state on the other hand features status heterogeneity, but lacks active links that connect nodes in different states and drive network dynamics: the network splits into several connected components, each of which is at absorbing consensus. Adaptive networks in active asymptotic states can feature dynamic equilibria (DEs) when state densities and
 other network motifs are constant in the thermodynamic limit, oscillatory behavior 
when motif densities are periodic functions of time,
 or more complex dynamical regimes (see \cite{Blasius2008,Gross2009} for reviews). Moreover for finite-size networks in the active phase, stochastic fluctuations induce transitions from a prolonged active to a frozen state. The lifetime distributions of these metastable regimes can be described as first-passage problems, and the scaling with system size of several relevant averages can be computed \cite{VazquezNJP2008,VazquezPRL2008,SoodPRL2005,Nardini2008}. Recent literature on adaptive networks focused on (different flavors of) two paradigmatic frameworks, each featuring individuals that cycle in binary state space and strive for homophilic interactions.

The first framework was proposed in \cite{VazquezPRL2008} and is the coevolutionary variant of the classic voter model (VM) \cite{Holley1975}, mimicking the symmetric spreading of two opinions. Its dynamics feed entirely on active links connecting the two competing node types A and B, and can be implemented in the following ways: In the \emph{node-update} scheme, at every time step a node \(i\) and one of its neighbors \(j\) are randomly picked. If the link connecting them is active then, in the direct (reverse) node-update, node i (node j) rewires it with probability \(w\) to another randomly selected node of the same type, and adopts the state of node \(j\) (node \(i\)) with probability \((1-w)\). If, instead, the link is inert, i.e. when it connects two nodes of the same type, then no action is taken. For the coevolutionary VM with \emph{link update}, a random link is picked. If it is inert, nothing happens. If it is active, one of its two end nodes is randomly chosen. Then that node rewires the active link with probability \(w\) to an arbitrarily picked node of the same type or adopts the state of the node at the other end of the link with probability \((1-w)\).

The ensuing dynamics are completely symmetric in node states, with the two update schemes yielding similar outcomes \cite{Demirel2013}, yet with one major difference. For link update, the average network magnetization - the difference in the two fractions of nodes types - is conserved. Under node update, network magnetization is conserved only when the network's degree distribution is sufficiently homogeneous \cite{SucheckiEPL2004}. Several modifications of the classic VM have been considered, aiming for more realistic representations of actual decision making in social contexts \cite{Galam2005, Galam2008}.

As the second framework, the contact process (CP) on an adaptive network put forward in \cite{GrossPRL2006} models the spreading of a disease in a population without immunity, but with disease awareness. Infected nodes (A-nodes) turn adjacent healthy nodes (B-nodes) into A-nodes with rate \((1-w)p\), while themselves recovering to B-nodes with rate \((1-w)(1-p)\) (\(w,p \in [0,1]\)). Additionally, healthy B-nodes evade infection by retracting links to infected neighbors with rate \(w\) and rewiring them to randomly selected B-nodes. Taking these three rate constants to add up to one amounts to a rescaling of system time that allows for the exploration of the model's phase diagram in a compact two-dimensional parameter space.  Unlike in the aforementioned VM, the elementary processes act on different network motifs: rewiring and infection occur along active links, whereas recovery operates on A-nodes regardless of their neighborhood composition.
 
The adaptive CP displays a rich dynamical behavior \cite{GrossPRL2006}, and its DEs have 
interesting properties. On one hand, the asymmetric dynamics generate strong correlations in the network due degree heterogeneity and node-state clustering, causing the system in DE to deviate from quantitative mean-field predictions. On the other hand, the corresponding metastable states in finite systems are in general long-lived, so that the DEs can be reliably sampled even for moderate system sizes.

In the following, we will consider a modification of the adaptive CP
that lets the recovery rate of infected nodes be modulated by the overall level of infection in the population. While the original model has a totally healthy population as the only possible consensus state, this modification allows for two dynamically competing consensus states and exhibits a rich phase diagram, see Sec.~\ref{s:mG}. Although difficult to motivate in the context of infection dynamics, this modification lends itself to a natural interpretation in terms of opinion dynamics, yielding a highly asymmetric extension of the coevolutionary VM with a biased rewiring rule and different update schemes for different node ensembles. The idea is that the two opposing opinions are associated with different social attitudes, which translate into different strategies to promote consensus in their holders' local environment. While biased voter dynamics have been put forward through different rates of opinion adoption  \cite{Williams1972} or interactions along directed links \cite{SanchezPRL2002,ZschalerPRE2012}, the asymmetric dynamics outlined here are a result of differing strategies for the two competing opinions.
In Secs.~\ref{s:models} and \ref{s:mG}, we detail the model and present its description in the pair approximation (PA). The overall performance of the PA is assessed in Sec.~\ref{s:comp} by comparison with Monte-Carlo (MC) simulations. In Secs.~\ref{s:pade} and \ref{s:mcde}, we focus on the active phase and its DE, and study the stochastic properties of the corresponding metastable state in finite-size networks. 
The main features of this model and of the standard coevolutionary VM are contrasted in Sec.~\ref{s:vm}. In Sec.~\ref{s:TP}, we give a closer look at a point in parameter space where the broken symmetry is partially restored by the choice of a particular combination for the rate constants of the competing 
node and link processes, and in Sec.~\ref{s:so} we conclude.
\section{The Model}\label{s:models}
In the proposed asymmetric coevolutionary opinion dynamics, B-nodes engage in the same dynamics as in the adaptive CP (and the link-update coevolutionary VM): Firstly, they promote homophily by retracting links from A-nodes with rate \(w\) and reattaching them to randomly selected B-nodes. Secondly, they adopt opinions from adjacent A-nodes with the transmission rate \((1-w)p\) (again \(w,p \in [0,1]\) without loss of generality). The third process in contrast describes opinion adoption in the A-ensemble and is a modification of the contact process' recovery rule: Randomly selected A-nodes (of fraction \(x\) of the total population) shall relax to B-nodes with a non-constant rate \((1-w)(1-p)(1+m)\), where the factor \((1+m)\) lets the network magnetization \(m\equiv (1-2x) \in [-1,1]\) steer the rate of the process.
These ensemble-specific update schemes reflect two competing strategies to promote consensus in a population: Segregationist B-nodes are orthodox opinion holders that spread their opinion via social pressure and strive for local consensus by seeking to interact with their peers. Proselytic A-nodes engage with and convert B-nodes in personal interactions, and their heterodox opinion relaxes to the B-ground state at a rate that reflects the overall dominance of that opinion.

\begin{figure}[ht]
\centering
\includegraphics[width=0.5\textwidth]{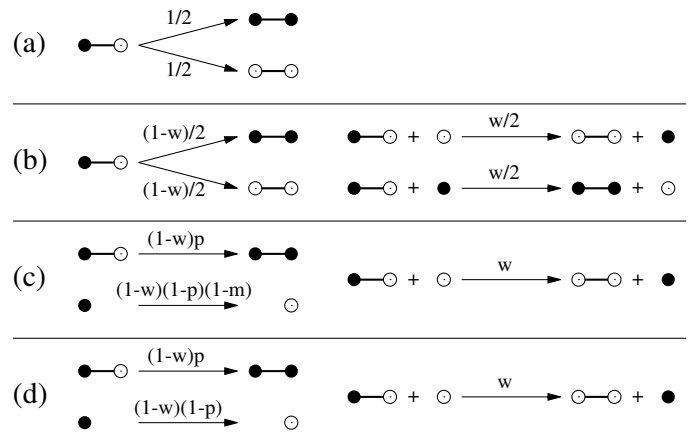}
\caption{Schematic representations of the \textbf{(a)} classic VM, \textbf{(b)} symmetric VM, \textbf{(c)} asymmetric VM and \textbf{(d)} adaptive CP, with filled (open) circles depicting A-nodes (B-nodes). Possible transitions between network motifs are illustrated by arrows, with the respective rates stated above.}
\label{f:models} 
\end{figure}

A schematic representation of this dynamics is depicted in Fig.~\ref{f:models}(c). It follows that while B-nodes engage in link-update coevolutionary voter dynamics,
the relaxation of A-nodes is boosted by a strong (global) presence of B-nodes, whereas it is diminished by a dominance of A-nodes. It can thus be seen as a mean-field description of the classic voter dynamics with (direct) node update, with the difference being that it is guided by overall network magnetization, not the magnetization of the neighborhood of the respective A-node. This global coupling on the part of A-nodes simplifies analytic modelling enormously and may also hint dynamical behavior for the latter case of coupling to the local magnetization.

A coarse-grained description of the network process is achieved by its moment expansion, where a set of evolution equations is derived for the per-capita density of network
motifs. In the subclass of pairwise models, only node and link densities are tracked, and the higher-order motif densities essential to describe their evolution are approximated.
This moment closure approximation caps the hierarchy of motif equations at the level of links, and it features a parameter \(\eta\) that translates the effect of the variance of the underlying network's degree distribution (see \cite{Demirel2013} for a review).

For the asymmetric opinion dynamics introduced above, the relevant link densities \(y\) and \(z\) refer to links among A-nodes and to active links connecting A- with B-nodes, respectively. Since rewiring ensures link-number conservation, the density of links connecting B-nodes is given by (\(\langle k \rangle/2-y-z\)) for \(0\leq (y+z) \leq\langle k \rangle/2\) and fixed mean degree \(\langle k \rangle\). The standard pair approximation (PA) of the process, in the spirit of \cite{GrossPRL2006}, then yields  

\begin{align}\label{e:pa}
\frac{dx}{dt}=&\left(1-w\right)\left(p \ z-\left(1-p\right)2(1-x) x\right)   \nonumber \\
\frac{dy}{dt}=&\left(1-w\right)\left(p \ z\left(\eta\frac{z}{1-x}+1\right)-2\left(1-p\right)2(1-x) y\right) \nonumber \\
\frac{dz}{dt}=&-z\left(w+\left(1-w\right)\left(p+\left(1-p\right)2(1-x)\right)\right)\nonumber \\&-\left(1-w\right)p\eta\frac{z^2}{1-x}\
+2\left(1-w\right)\left(1-p\right)2(1-x) y \nonumber \\ 
&+2\left(1-w\right)p\eta\frac{\left(\langle k \rangle -y-z\right)z}{1-x} \,.
\end{align}
The last term of the time evolution of \(z\) in Eqs.~(\ref{e:pa}) for instance describes the gain in active links via the adoption of the A-state through either end of a link connecting two B-nodes. In that case, the density of the relevant triplet motif (consisting of a central B-node connected to both another B- and an A-node) is approximated using the densities of the aforementioned link types and of B-nodes. 

In the dynamics described by Eqs.~(\ref{e:pa}) (referred to as \emph{asymmetric VM} in the following), link aquisition and opinion adoption are tailored to specific node ensembles, so that highly-skewed degree distributions can ensue for a wide range of parameters (see \cite{GrossPRL2006, WielandEPJ2012} as examples for related dynamics). For that reason, instead of the regular random graphs taken in \cite{VazquezNJP2008,VazquezPRL2008}, we decide for initial  Erd\H{o}s-R\'{e}nyi (ER) graphs (featuring a wider Poissonian degree distribution) and keep \(\eta=1\) throughout, for it has been shown that for even more heterogenous degree distributions, the respective moment closure maintains validity \cite{GrossPRL2006,Demirel2013}. As we shall see, with that choice of \(\eta\) the PA gives a good quantitative description of the corresponding network process. Moreover, we will focus on the long-term behavior of the dynamics and compare it to the asymptotic scenarios of the coevolutionary VM in \cite{VazquezPRL2008} (in the following referred to as \emph{symmetric VM} if topological coevolution is featured, and as \emph{classic VM} otherwise). The elementary processes defining all previously introduced dynamics are summarized in Fig.~\ref{f:models}.

\section{Asymptotic States in the PA}\label{s:mG}
Denoting the state vector of Eqs.~(\ref{e:pa}) as \((x,y,z)\) with \(x \in [0,1]\), \(y \in [0,\langle k \rangle /2]\), and \(z \in [0,\langle k \rangle /2-y]\) yields the two A- and B-consensus states \((0,0,0)\) and \((1,\langle k \rangle /2,0)\), respectively. It is straightforward to check that these are equilibria for Eqs.~(\ref{e:pa}). A linear stability analysis of the PA reveals that the B-consensus is stable for 
\[p\leq\frac{2-w}{\left(2+\langle k\rangle\right)\left(1-w\right)}.\]

As the PA equations are singular at \((1,\langle k \rangle /2,0)\), one needs to resort to regularization techniques to obtain
\[p>\min\left(\frac{2}{1+\langle k \rangle},\frac{2-3w}{1-w}\right) \]
as the parameter region for stable A-consensus (see Sec.~\ref{s:app}).

\begin{figure}[ht]
\centering
  \includegraphics[width=0.5\textwidth]{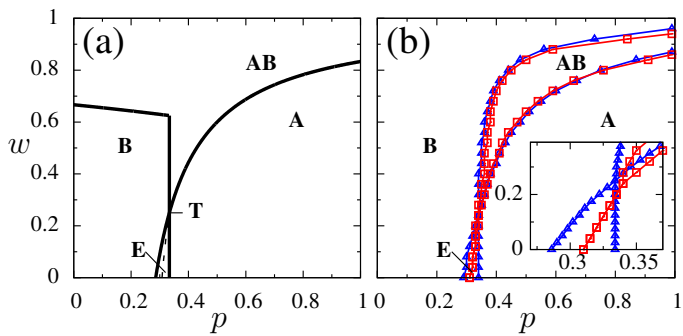}
  \caption{(Color online) \textbf{(a)} Phase diagram with steady states A-consensus (\(A\)), B-consensus (\(B\)) and the active phase (\(E\)), as well as the triple point (\(T\)), bounded by solid lines. A sequence of symbols indicates the coexistence of respective attractors. The dashed line \(x_{\rm E}=1/2\) yields DEs of the adaptive CP for respective parameter values. \textbf{(b)} Change of asymptotic behavior in the PA with initial conditions \((0.1,0.025,0.45)\), \((0.5,0.625,1.25)\), and \((0.9,1.025,0.45)\) (numerical integration of PA, blue triangles) from initially connected ER graphs with fractions \(0.1\), \(0.5\), and \(0.9\) of randomly assigned A-states (MC simulations, red squares). Regions of equal asymptotic behavior are marked by the same sequence of symbols as in (a). For sufficiently long simulation runs, stochastic fluctuations in the metastable state drive the full system into A- or B-consensus [inset of (b)].  Mean degree \(\langle k \rangle=5\), MC simulations with \(N=5000\) nodes and results averaged over 100 realizations.}
\label{f:mG} 
\end{figure}

Apart from a frozen phase, the PA yields a DE for a small parameter region
\begin{equation}\label{e:xderange}
\frac{2-w}{\left(2+\langle k\rangle\right)\left(1-w\right)}\leq p\leq\frac{2}{1+\langle k \rangle}
\end{equation}
bordering the two consensus states. In this region of intermediate \(p\) and small \(w\), relaxation and transmission balance out, with the small rewiring rate allowing for the continued existence of a nonzero density of active links and a steady-state fraction of A-nodes of
\begin{equation}\label{e:xa}
x_{\rm E}=\frac{2-(2+\langle k\rangle)p(1-w)-w}{w-p(1-w)}\,.
\end{equation}
The system's only active phase ensues, whose size in the phase diagram shown in Fig.~\ref{f:mG}(a) decreases for increasing mean degree \(\langle k \rangle\). For \(x_{\rm E}=1/2\), the magnetization of the system in steady state is zero while still in the active phase, so that A-nodes recover with constant rate \(r\) and the PA describes the DE of the adaptive CP at the respective parameters.

The triple point \(T\) in Fig.~\ref{f:mG}(a) at
\begin{align*}
p&=\frac{2}{1+\langle k \rangle}\equiv p_{\rm T}\\
w&=\frac{2}{3+\langle k \rangle}\equiv w_{\rm T}
\end{align*}
lies at the confluence of the active and the two absorbing phases. It marks the end of the active phase and the switch for faster rewiring to a bistable regime of coexisting consensus states. In the bistable regime at a given rewiring rate \(w\), the parameter \(p\) tunes the competition between the two consensus states through the size of their basins of attraction.

The subensemble mean degrees in DE are given by the PA as
\begin{align*}
\langle k_\mathrm{A} \rangle&=\frac{2}{p}-1 \\
\langle k_\mathrm{B} \rangle&=\frac{2-2p(1-w)-w}{p(1-w)} \, ,
\end{align*}
so that
\begin{equation}\label{e:vm1Ksi}
\langle k_\mathrm{B}\rangle -\langle k_\mathrm{A}\rangle=\frac{w}{(1-w)p}-1 \, ,
\end{equation}
similarly to findings for the DE in the adaptive CP \cite{WielandEPJ2012}. Throughout the active phase, \(p\leq w/(1-w)\) holds, with the equality fulfilled only at the triple point. It follows that in steady state, the mean degree of the A-ensemble is larger than that of the B-ensemble, despite the rewiring bias towards B-nodes [see also Fig. \ref{f:T}(b)].

\section{Comparison to the Stochastic Network Process}\label{s:comp}

To properly compare the PA with MC simulations, one has to faithfully translate network configurations into PA state vectors.
An ER graph with a fraction \(x_0\) of randomly primed A-nodes is described by the PA as \((x_0,\langle k \rangle x_0^2/2,\langle k \rangle (1-x_0)x_0)\). Initial conditions in all MC runs are set this way and translated into the PA formalism accordingly. The simulations are implemented following \cite{Gillespie1976}. To emulate networks with \(N\) nodes in the PA, node densities smaller than \(1/N\) and larger than \(1 - 1/N\) are taken to represent the two consensus states in a network of that size, and integration of Eqs.~(\ref{e:pa}) is stopped as soon as \(x\) reaches any of these two values. 

Furthermore, it is important to identify "pathological" network configurations and avoid them altogether. If for instance MC simulations without rewiring ran on networks with isolated B-subgraphs (consisting of only B-nodes), A-consensus could not be reached, as these B-subgraphs would be left unchanged by network dynamics. This would moreover add a
constant offset to the network's magnetization that would distort relaxation in the remaining components of the network.

As \(\langle k \rangle<\log(N)\) for mean degrees \(\langle k \rangle\) and system sizes \(N\) used in MC simulations here, an initial ER graph is almost surely fragmented \cite{Erdos1960}. Its isolated subgraphs need to be linked through i) linking two randomly selected nodes from separate subgraphs ii) randomly picking a node that emanates links of the same type added in i), randomly choosing and deleting one of them iii) repeating i)-ii) until the graph is connected. Because the initial ER graph has a Poissonian degree distribution, a lower limit on its number of disconnected components can be given through  \(e^{-\langle k\rangle }N\), where \(e^{-\langle k\rangle}\) is the average fraction of isolated nodes. For the values of \(\langle k \rangle\) and \(N\) used, this lower limit approximates the actual number of initially disconnected subgraphs very well. Thus the fraction of nodes involved in this linking procedure is very small. The procedure introduces no correlations either in degree or in status, so that apart from the vanishing isolated nodes, the main characteristics of an ER network are preserved. All initial networks used in the following MC simulations are connected this way.

When stable asymptotic states coexist, we have to take into account their basins of attraction. Selecting a sufficiently large set of initial conditions and monitoring the resulting asymptotic behavior of the system allows for the detection of all basins of attraction, both in integration of Eqs.~(\ref{e:pa}) and MC simulations. Browsing parameter space with this procedure would lead to a comparison of the phase boundaries of Fig.~\ref{f:mG}(a) with their MC analogue. 
Instead, we identify regions in parameter space for which a given small set of initial conditions lets dynamics drive the PA and the full system into the same set of asymptotic states. Comparing such parameter regions resulting from integration of Eqs.~(\ref{e:pa}) to those obtained from MC simulations allows for a quantitative comparison of PA dynamics and the corresponding network process without having to verify phase boundaries of Fig.~\ref{f:mG}(a). 

This coarse-grained browsing of initial conditions over the whole parameter space yields a good agreement between numerical integration of the PA and MC simulations [Fig.~\ref{f:mG}(b)], indicating that for initial (connected) ER graphs, the PA faithfully models the actual dynamics. An \emph{AB}-parameter region in that context means that either A- or B-consensus can be reached from the set of initial conditions used, whereas regions \emph{A}, \emph{B} and \emph{E} signal a uniform asymptotic behavior leading to A-consensus, B-consensus, and a DE, respectively. In MC simulations, a metastable DE is observed for parameter values of the PA's active phase. In it, stochastic fluctuations eventually drive the system into one of the two consensus states. A more thorough PA description of the active phase, as well as a stochastic modeling of the corresponding metastable DE in the network process, will be given in Secs.~\ref{s:pade} and \ref{s:mcde}.
\section{The Active Phase in the PA}\label{s:pade}
\begin{figure}[ht]
\centering
  \includegraphics[width=0.5\textwidth]{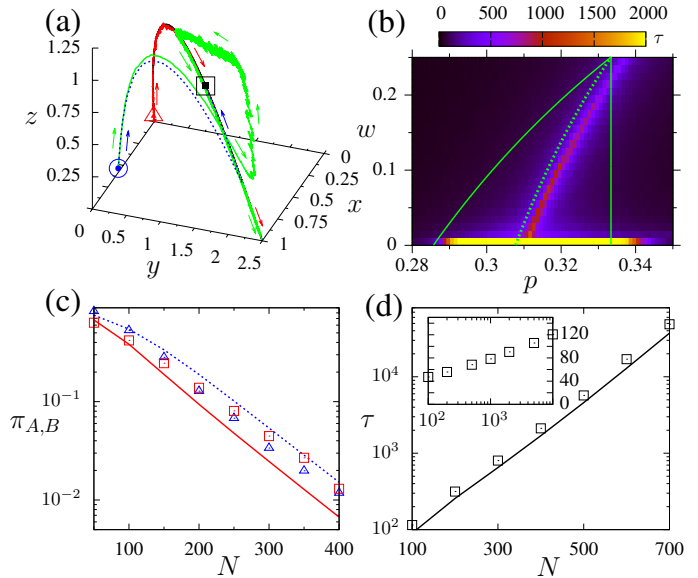}
\caption{(Color online) Active phase in the asymmetric VM. \textbf{(a)} Time evolution of \(x\), \(y\) and \(z\) for \(w=0.05\) and \(p=0.32\) along \(M_{\rm E}\) (solid black line connecting consensus states), with the DE on \(M_{\rm E}\) marked by a square and arrows indicating the respective trajectory's direction. All MC trajectories align with \(M_{\rm E}\), pass through the DE and end up in A-consensus, starting from \((0.01,0.00025,0.0495)\) (triangle corresponding to an initial ER graph; solid red line) and \((0.8,0.2,0.2)\) (circle corresponding to a maximally random graph with respect to initial conditions, solid green line overshooting and eventually aligning with \(M_{\rm E}\)). The latter initial network is obtained through \textbf{i)} generating two separate random A- and B-subgraphs compatible with given \(x\), \(y\), and \((\langle k \rangle/2-y-z)\) as well as \textbf{ii)} connecting them through \(N z\) randomly assigned active links. Numerical integration from \((0.8,0.2,0.2)\) (blue dashed line) ends up in DE at \((0.701,1.394,0.891)\) (square). Network size \(N=10^4\) in MC runs. \textbf{(b)} Color-coded convergence times \(\tau\) in MC simulations within DE phase boundaries obtained from PA (solid green lines); maximum \(\tau\) are expected at \(x_{\rm E}=1/2\) (dashed green line). MC simulations with \(N=5000\), averaged over \(100\) runs. \textbf{(c)} System-size dependent splitting probabilities for \(w=0\) computed through Eq.~(\ref{e:pi}) (lines) and fixation probabilities taken from MC simulations (symbols). \(\pi_{\rm A}\) is computed for \(p=0.305\) starting from \(x_0=0.9 \) (blue triangles and dashed line) and \(\pi_{\rm B}\) for \(p=0.315\) starting from \(x_0=0.1\) (red squares and solid line). \textbf{(d)} Convergence times in MC simulations (squares) and from Eq.~(\ref{e:tautau}) as a function of system size for \(w=0\) and \(p=0.3\). Inset: MC simulations for \(w=0.05\). MC simulations averaged over \(10^4\) runs (\(10^2 \leq N <10^3\)) and \(10^3\) runs (\(10^3 \leq N <10^4\)). Mean degree \(\langle k \rangle=5\) in all figures. MC simulations in (b)-(d) from initially connected ER graphs, in (b) and (d) with \(x_0=0.5\).}
\label{f:de} 
\end{figure}

For Eqs.~(\ref{e:pa}) in the active phase, the transient dynamics are reminiscent of what is reported from the symmetric VM \cite{VazquezPRL2008}, in that the deterministic system relaxes quickly to a parabola-shaped slow manifold \(M_{\rm D}\) [Fig.~\ref{f:de}(a)].  In our case on the other hand, \(M_{\rm D}\) is generally not a line of equilibria, but spanned by two heteroclinic orbits connecting the stable node (the DE) with the two saddles that represent the unstable consensus states. Once driven to \(M_{\rm D}\), the system moves slowly towards the DE along one heteroclinic orbit. Strictly speaking, \(M_{\rm D}\) cannot be classified as a slow manifold, as the latter is associated with a degenerate eigenvalue of the linearized flow, while the DE is linearly stable throughout the active phase (except at the triple point \(T\)). In the following however, we widen the definition to any set of trajectories that the flow quickly relaxes to and then slowly proceeds along towards an attracting fixed point.

The slow manifold \(M_{\rm D}\) is well approximated by a curve \(M_{\rm E}\) given by
\begin{align}\label{e:MA}
 M_{\rm E}&=\begin{pmatrix} x  \\y_{\rm E}\{x\} \\z_{\rm E}\{x\} \end{pmatrix}\nonumber \\
&=\begin{pmatrix} x  \\x\frac{2 x (x-\langle k\rangle)-2+w (1+(3+2 \langle k\rangle-4 x) x)}{2 (w+w x-2)} \\2 (1-x) x\frac{ \langle k\rangle (w-1)+w+x-2 w x}{w+w x-2} \end{pmatrix} \ ,
\end{align}
with \(y_{\rm E}\{x\}\) and \(z_{\rm E}\{x\}\) being the equilibrium values of link densities \(y\) and \(z\) for an A-node fraction settling down to \(x\). It follows that \(M_{\rm E}\) is the set of all DEs that, for fixed \(\langle k \rangle\) and \(w \leq w_{\rm T}\), are generated by all \(p\) for which the system is in the active phase, given by the interval in Eq.~(\ref{e:xderange}).

As \(p\) enters this interval from smaller values that lead to B-consensus, a transcritical bifurcation turns the stable node in \((0,0,0)\) into a saddle, emanating a stable DE that is moving along \(M_{\rm E}\). It reaches \((1,\langle k \rangle/2,0)\) at \(p=p_{\rm T}\) and, in another transcritical bifurcation, vanishes while turning the saddle there into a stable node representing A-consensus. 

Since the vector field of the PA along \(M_{\rm E}\) is generally not tangent to the latter, \(M_{\rm E}\) is usually  not a trajectory of the system and hence does not coincide exactly with  \(M_{\rm D}\). It is however straightforward to show that this matching improves for increasing \(\langle k\rangle\) and \(w\), while it is already very good for the low mean degrees and rewiring rates considered here. As a first approximation, the description of the latter stages of system evolution towards the DE can, as in \cite{VazquezPRL2008,VazquezNJP2008}, consequently be collapsed to one variable \(x\), constraining the remaining two to be on \(M_{\rm E}\).

At the triple point \(T\), i.e., for the highest rewiring rate still allowed in the active phase, the range of \(p\) for which there is an active phase shrinks to the single value \(p=p_{\rm T}\) [Eq.~(\ref{e:xderange}) and Fig.~\ref{f:mG}(a)]. It can be shown that then \(M_{\rm E}\) and \(M_{\rm D}\) exactly coincide, with the PA yielding a continuum of transversally stable stationary states given by Eq.~(\ref{e:MA}). A more thorough description of the model phenomenology at \(T\) will be given in Sec.~\ref{s:TP}.
	
\section{The Active Phase in the Full System}\label{s:mcde}
The onset of the metastable DE in MC simulations is characterized by vastly increasing convergence times needed to reach a consensus state, particularly for zero rewiring [Fig.~\ref{f:de}(b)]. Projecting the network's time evolution onto the reduced phase space spanned by \((x,y,z)\), the resulting random walk (RW) of the stochastic system follows closely the trajectory of its PA description along a curve \(M_{\rm S}\) [Fig.~\ref{f:de}(a)], but with three fundamental differences:

1) The exact shape of \(M_{\rm S}\)  depends on the system size \(N\) and approaches \(M_{\rm D}\) (the slow manifold of the PA) as \(N\) increases.

2) For various starting conditions well apart from \(M_{\rm S}\) [such as those marked by a circle in Fig.~\ref{f:de}(a)], the RW does not immediately relax to \(M_{\rm S}\). Instead, the network undergoes a sequence of distinct configurations before it realigns with the reduced description given by \(M_{\rm E}\). These intermediate network configurations are dependent on initial conditions and partition the RW into several segments.

3) Stochastic fluctuations drive the system along \(M_{\rm S}\) from the DE towards one of the two consensus states. Unlike in the stochastic process in the symmetric VM, the type of this final consensus state in network dynamics is, for sufficiently large system sizes, predetermined by the system's parameters and independent of initial conditions (see below).  

Following the approach successfully carried out for the symmetric VM \cite{VazquezNJP2008,VazquezPRL2008}, we will explore the possibility of reducing the description of asymmetric dynamics in the full system to a RW along \(M_{\rm S}\), while considering the simplest case of zero rewiring. Lacking an analytic expression both for \(M_{\rm S}\) and \(M_{\rm D}\), the RW is assumed to take place along the approximate slow manifold \(M_{\rm E}\) given by Eq.~(\ref{e:MA}), yielding a one-step process in the total number \(X\) of A-nodes with the master equation 
\begin{align}\label{e:me}
\frac{\partial [X]}{\partial t}=&p  z_{\rm E}\{X-1\}[X-1]\nonumber \\
&+2(1-p)\left(1-\frac{X+1}{N}\right)(X+1)[X+1] \nonumber \\
& -p  z_{\rm E}\{X\}-2(1-p)\left(1-\frac{X}{N}\right)X[X]. 
\end{align}

Here \([X]\) is the time-dependent probability for a network of size \(N\) to have \(X=N x\) A-nodes, and \(z_{\rm E}\{X\}=N z_{\rm E}\{x\}\) is the total number of active links computed from the respective link density at \(w=0\) in Eq.~(\ref{e:MA}). The consensus states \(X=0\) and \(X=N\) are absorbing boundaries of the RW with starting point \(X_0 \in [0,N]\), so that the eventual termination of the metastable state in the full system corresponds to a first-passage problem. 
\subsection{Fixation probabilities}\label{s:pa}
One is interested in the splitting probability \(\pi_{\rm A}\{X_0,N\}\) (resp. \(\pi_{\rm B}\{X_0,N\} = 1-\pi_{\rm A}\{X_0,N\}\)) for the RW to start at \(X=X_0\) and end up at \(X=N\) (resp. \(X=0\)), approximating the fixation probability for A-consensus (resp. B-consensus) in the full system. With Eq.~(\ref{e:me}) and following \cite{vanKampen2007}, we obtain for \(w=0\)
\begin{equation}\label{e:pi}
\pi_{\rm A}\{X_0,N\}=\left(1+\frac{\sum_{K=X_0}^{N-1} P(K)}{1+\sum_{K=1}^{X_0-1} P(K)}\right)^{-1}
\end{equation}
(see Appendix \ref{s:app2}), where 
\begin{equation*}
P(K) = \prod\limits_{X=1}^K \frac{2(1-p)(1-X/N)X}{p  z_{\rm E}\{X\}} \, .
\end{equation*}

The behavior of \(\pi_{\rm A, \rm B}\{X_0,N\}\) for increasing \(N\) and constant \(x_0=X_0/N\) is given by Eq.~(\ref{e:pi}) and shown in Fig.~\ref{f:de}(c) for two different values of \(p\) that illustrate the case when the DE is closer to B-consensus than to A-consensus (\(p = 0.305\)) and the opposite case (\(p = 0.315\)). In each case, the starting point \(X_0\) is taken close to the consensus state that is farthest from the DE. Equation~(\ref{e:pi}) predicts that even for moderate system sizes, the final consensus type will, with overwhelming probability, be the one that is closest to the DE [dashed blue and solid red line in Fig.~\ref{f:de}(c)]. Monte-Carlo simulations also shown in Fig.~\ref{f:de}(c) (blue triangles and red squares) confirm that, as system size grows larger, the final consensus state is increasingly determined by the position of the DE on the slow manifold. This is plausible considering that for sufficiently large \(N\), the system can be seen as initially drifting towards the DE, and from there diffusing to consensus. That also applies to \(w>0\), so that, as \(N\) becomes larger, stochastic fluctuations generally drive the system from the metastable to a consensus state whose type is increasingly independent of initial conditions. The fixation probability for the consensus state
farthest from the DE is observed to decrease exponentially with \(N\) for large enough \(N\), a scaling corroborated by Eq.~(\ref{e:pi}).

The convergence to \(1\) or \(0\) of the fixation probabilities \(\pi_{\rm A, \rm B}\{X_0,N\}\) is in contrast with the result \(\pi_{\rm A}\{X_0,N\}=x_0\) that holds for the symmetric VM. Equation~(\ref{e:pi}) yields this result in the particular case when the factors in the products \(P(K)\) are identically equal to 1, that is, when  
\begin{equation*}\label{e:add}
p  z_{\rm E}\{X\}=2(1-p)(1-X/N)X
\end{equation*}
at all \(X\). This would correspond to the existence of a line of equilibria along which  transmission events and relaxation events balance out. 

In general, the heuristic rule holds that the final consensus state for large \(N\) is the one whose distance to the DE, to be bridged by stochastic fluctuations, is smaller.
This rule is corroborated by Eq.~(\ref{e:pi}) that, in the exemplary  case of \(\langle k \rangle=5\) and \(w=0\) in Fig.~\ref{f:de}(b), yields \(p=0.308\) as the value for which 
the most likely final consensus state switches
when varying \(p\), with the \(x\)-coordinate of the DE for that value of \(p\) being \(x_{\rm E}\{0,0.308,5\}=0.509\). For \(x_{\rm E}=1/2\) then, i.e., when the PA describes the adaptive CP in steady state, one would conversely expect maximum first-passage times for the RW along the slow manifold, and therefore also maximum convergence times \(\tau\) for the full system. Indeed this assumption yields a good estimate for the \(p\)-coordinate of maximum \(\tau\) both for zero rewiring and for \(w \neq 0\). The line of maximum convergence times in Fig.~\ref{f:de}(b) differs slightly from \(x_{\rm E}(w,p,\langle k \rangle)=1/2\) because the PA does not give the exact DE coordinates of the full system. Using in contrast the \(X\)-coordinates of the DE in MC simulations yields a very good match.

Except for the qualitative considerations above, computing \(\pi_{\rm A}\{X_0,N\}\) with Eq.~(\ref{e:pi}) only yields moderately accurate predictions for the full system without rewiring [Fig.~\ref{f:de}(c)], even when using \(M_{\rm S}\) (as sampled from an ensemble of trajectories in corresponding MC simulations) instead of the approximate \(M_{\rm E}\). The reason is that since convergence times diverge quickly on static networks [see below and Fig.~\ref{f:de}(d)], the computation of \(\pi_{\rm A}\{X_0,N\}\) is only feasible for relatively small system sizes. Yet for small \(N\), assuming that the stochastic dynamics take place along a smooth \(M_{\rm S}\) neglects the effect of significant transversal fluctuations: Computing \(\pi_{\rm A}\{X_0,N\}\) along an averaged \(M_{\rm S}\) (taken over many stochastic trajectories) generally yields different results than directly averaging \(\pi_{\rm A}\{X_0,N\}\) over the ensemble of trajectories. This remains an issue even for increasing system sizes, as \(M_{\rm S}\) retains "problematic" segments of low densities \(x\) and or \(z\) in the vicinity of the two consensus states, where transversal fluctuations are large even for large \(N\).

These limitations do not apply to the symmetric dynamics along a line of equilibria considered in \cite{VazquezNJP2008,VazquezPRL2008}, because in that case transversal fluctuations are decoupled from the random walk in X. In our asymmetric VM however, a stable DE is present on \(M_{\rm S}\) with both the drift and diffusion of the random walk in \(X\) depending on the link density \(z\). Even if \(M_{\rm S}\) were a line of equilibria 
(Sec.~\ref{s:vm}), transmission and relaxation in its vicinity would generally not balance out, so that the (finite) full system with transversal fluctuations is not captured by its reduced description along \(M_{\rm S}\). Consequently, a new framework is needed to address metastability in the asymmetric VM.
\subsection{Convergence times}\label{s:va}
Similarly to Sec.~\ref{s:pa}, one can derive an expression for 
\begin{equation*}
\nu_{\rm A,B}\{X_0,N\}\equiv\pi_{\rm A,B}\{X_0,N\}\tau_{\rm A,B}\{X_0,N\} \, ,
\end{equation*}
where \(\tau_{\rm A,B}\{X_0,N\}\) denotes for each consensus state the mean first-passage times, with \(X_0\) as starting point, of the process in Eq.~(\ref{e:me}). Following Appendix \ref{s:app3} and using the transition rates for this process, one obtains an analytic expression for the mean time to achieve consensus when starting from \(X_0\), the \emph{convergence time}
\begin{equation}\label{e:tautau}
\tau\{X_0,N\}=\nu_{\rm A}\{X_0,N\}+\nu_{\rm B}\{X_0,N\} \, .
\end{equation}
Setting  \(x_0 = 1/2\), \(\tau\{X_0,N\}\) as a function of \(N\) given by Eq.~(\ref{e:tautau}) is plotted in Fig.~\ref{f:de}(d) together with the results of MC simulations for \(w=0\) and the same starting point. 
It can be seen that the analytic approximation of Eq.~(\ref{e:tautau}), based on the reduction of the full stochastic system to a RW along the slow manifold, yields a good quantitative agreement with convergence times observed in the full system.

For  \(w=0\), we obtain exponential scaling of convergence times with system size, as found in another modification of the symmetric VM \cite{Rogers2013} also featuring a slow manifold connecting a stable heterogeneous state and unstable consensus. This is in stark contrast with the scaling of convergence times \(\tau\) with system size \(N\) for \(w \neq 0\), which appears to be sublinear [inset of Fig.~\ref{f:de}(d)]. The latter scaling may seem surprising given that in both situations, there is a drift towards a stable DE on the slow manifold countering diffusion towards consensus. 
A similar dramatic effect of rewiring on the scaling properties of convergence times occurs for the symmetric VM. As described and explained in \cite{Nardini2008}, in the symmetric VM consensus is strongly favored or disfavored by rewiring according to whether the direct or reverse update scheme is adopted, due to small changes of the flow on and close to the slow manifold. By contrast, in the asymmetric VM, rewiring always favors consensus. The explanation must be found in the drift rates along the slow manifold becoming smaller for \(w \neq 0\) while rewiring increases overall degree heterogeneity, leading to enhanced fluctuations.
\section{Comparison to Symmetric Coevolutionary Voter Dynamics}\label{s:vm}
In the symmetric VM of \cite{VazquezPRL2008}, the network magnetization \(m\) is conserved in the thermodynamic limit for sufficiently homogeneous initial  graphs \cite{SucheckiEPL2004}, and a single ODE suffices to describe the full system: It yields a (frozen) fragmented phase and an active phase which features a continuum of stable steady states. The specific steady state to be reached is then determined by the parameter \(m\). For finite system sizes and due to stochastic fluctuations, the corresponding metastable state decays into either consensus state, following closely a slow manifold formed by a continuum of steady states. It does so along a slow manifold given by aforementioned continuum of steady states. Consequently \(\pi_{\rm A}\{X_0,N\}=X_0/N\), that is, the fixation probabilities are system-size independent and depend linearly on the starting coordinates \(X_0\) on the slow manifold. Therefore consensus in the symmetric VM is reached from a metastable DE in the active phase, that is, only stochastically, not dynamically. The model's symmetry moreover precludes any coexistence of attractors; however \emph{stochastic} bistability of the consensus states in the metastable state is given by the system-size invariant fixation probabilities above.   

In contrast, the asymmetric VM proposed here lets two voter-model update schemes compete against each other, breaking the node-ensemble symmetry in a twofold way: Through state-dependent imitation rules and through tying the rewiring rule to just one specific node ensemble. The broken symmetry demands for additional degrees of freedom to capture the full system, yielding Eqs.~(\ref{e:pa}). As a consequence, network magnetization is not conserved anymore, but a system variable that guides relaxation. This in turn impedes a frozen state with \(0<x<1\), so that (by construction) the asymmetric VM lacks a fragmented phase. The slow manifold is generally not a line of equilibria anymore, since isolated equilibrium points - one in each consensus state and, in the active phase, an additional stable DE - replace the continuum of steady states. It follows that i) the steady state in the active phase is now independent of initial conditions, particularly \(m\) ii) consensus can now be reached dynamically iii) consensus bistability is predicted by the PA and observed for the full system iv) for consensus reached stochastically in the metastable state, the fixation probabilities are now system-size dependent, converging for large system sizes to either zero or one regardless of initial conditions.

Unlike the symmetric VM, the asymmetric VM is not robust against varying topological backgrounds. When choosing for instance \(\eta=(\langle k \rangle-1)/ \langle k \rangle\) in Eqs.~(\ref{e:pa}) to assume an initial random regular graph \cite{Demirel2013}, the intersection of the two consensus boundaries in phase diagram Fig.~\ref{f:mG}(a) disappears, and with it the active phase. Instead, bistable consensus stretches down to \(w=0\), featuring a slow manifold much like the active phase of the model with \(\eta=1\) did (see Sec.~\ref{s:pade}). This time however it is spanned by heteroclinic orbits connecting two \emph{stable} nodes in the consensus states with a saddle as the \emph{unstable} DE. The different model phenomenology for that choice of \(\eta\) is corroborated by MC simulations, emphasizing the importance of initial topology for asymmetric opinion dynamics on adaptive networks. Since changing network topologies may shift the balance in opinion competition, the initial topology can be crucial for the outcome of the asymmetric dynamics \cite{AntalPRL2006}, and this applies even to very large systems because transient duration increases with system size.
The (also asymmetric) adaptive CP on the other hand generally allows for sufficiently enduring coevolutionary dynamics to wash out initial differences in network structure. This is because its active phase does not feature a slow manifold along which stochastic fluctuations towards consensus could be facilitated. 
\section{The triple point}\label{s:TP}
Tuning the rates of the asymmetric VM's three elementary processes strengthens or loosens its asymmetry.
For the parameters \((w_{\rm T},p_{\rm T})\) defining the triple point \(T\) that borders all phases, 
the PA in steady-state is exactly captured by the slow manifold in Eq.~(\ref{e:MA}) referred to as \(M^{\rm T}_{\rm E}\) (Sec.~\ref{s:pade}), so that
\begin{itemize}
\item  \(M^{\rm T}_{\rm E}\) is formed by equilibrium points, as transmission and relaxation events balance out according to \(p  z_{\rm E}\{x\}=2(1-p)(1-x)x\).
\item \(\langle k_{\rm A}\rangle=\langle k_{\rm B}\rangle =\langle k \rangle\) for mean degrees \(\langle k_{\rm A}\rangle=(2y+z)/x\) and \(\langle k_{\rm B}\rangle=(2(\langle k\rangle/2-y-z)+z)/(1-x)\) of the A- and B-ensemble, respectively.
\end{itemize}
These two steady-state equalities are not found anywhere else in the active phase of the asymmetric VM and moreover yield \(z_{\rm E}\{1-x\}=z_{\rm E}\{x\}\) as well as \(y_{\rm E}\{1-x\}=\langle k \rangle/2-y_{\rm E}\{x\}-z_{\rm E}\{x\}\) as a noteworthy consequence, i.e., at \(T\) the system stays in DE even if all node states are flipped, as it does in the symmetric VM.

It is straightforward to show that the line of equilibria given by \(M^{\rm T}_{\rm E}\) is the same as for the PA of the symmetric VM with link-update and \emph{without} rewiring. Therefore at \(T\), that is, for a nontrivial choice of parameters, the steady states of the asymmetric VM replicate the active phase of  (one particular flavor of)  the classic VM. 

An additional feature unique to the triple point in the asymmetric VM contrasts topology change with opinion spreading: at \(T\), transmission along and rewiring of active links happen at the same \emph{rate} \((1-w)p=w\). Combined with the balance of transmission and relaxation \emph{events}, this implies the equipartition of processes in steady state at \(T\): Rewiring, relaxation and transmission then each account for exactly one third of events. As soon as rewiring dominates through \(w>w_{\rm T}\), topology change impedes a dynamic equilibrium [Fig.~\ref{f:mG}(a)].

For the symmetric VM with any rewiring rate, describing metastability of the DE with a one-step process in \(X=Nx\) as in Sec.~\ref{s:mcde} gives excellent quantitative results. This is due to the fact that in this case the flow conserves \(x\) throughout - on and off the slow manifold - so that fluctuations transversal to the slow manifold relax with no effect on the RW in \(X\).

At the triple point, the slow manifold of the asymmetric VM is drift-free, as in the symmetric VM.
However, the respective flows are fundamentally different even in the vicinity of the line, and for the full system at \(T\), transversal fluctuations of a trajectory along the slow manifold should still experience a coordinate-dependent drift in the \(x\)-direction. Because of that, even at \(T\) the ansatz of Sec.~\ref{s:mcde} to characterize metastability of DEs is of limited use.

\begin{figure}[ht]
\centering
  \includegraphics[width=0.5\textwidth]{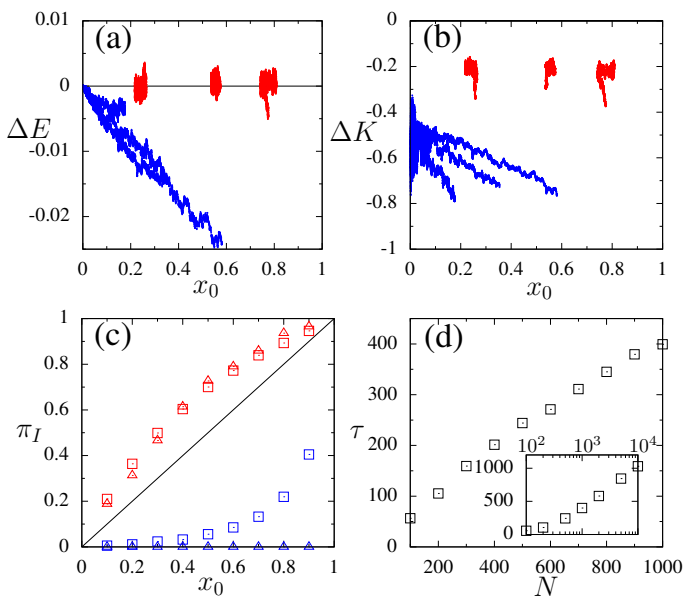}
\caption{(Color online) Identifying the triple point in the full system with plots for \((w_{\rm T}^*,p_{\rm T}^*)\) and \((w=0.05,p=0.3)\) dynamically [(a)-(b)] and stochastically [(c)-(d)]. \textbf{(a)} Balance of events \(\Delta E =p \ z-2(1-p)(1-x)x\) and \textbf{(b)} of mean degrees \(\Delta K=\langle k_{\rm B}\rangle-\langle k_{\rm A}\rangle\) for simulation runs from fractions \(x_0=0.2\), \(0.5\), and \(0.8\) of randomly primed A-nodes. The relatively stationary trajectories (red, upper half of both figures) correspond to \((w_{\rm T}^*,p_{\rm T}^*)\), whereas the ones approaching B-consensus (blue, lower half) were recorded for \((w=0.05,p=0.3)\). \textbf{(c)} Fixation probability for A-consensus as a function of \(x_0\) for \(N=100\) (squares) and \(N=1000\) (triangles), with red data points above the diagonal line obtained for \((w_{\rm T}^*,p_{\rm T}^*)\) and blue data points below for \((w=0.05,p=0.3)\). \textbf{(d)} Scaling of convergence times with system size at \(T^*\). MC simulations in (a) and (b) recorded for \(N=10^5\) and \(10\leq t\leq 100\), in (c) and (d) averaged over \(10^3\) runs. All simulations from initially connected ER graphs and with mean degree \(\langle k \rangle=5\).}
\label{f:T} 
\end{figure}

In MC simulations, we find that for the approximate coordinates \((w_{T}^*\approx0.150, p_{T}^*\approx0.325)\) instead of \((w_{\rm T}=1/4,p_{\rm T}=1/3)\), the full system with \(\langle k \rangle =5\) features a drift-free attracting curve \(M_{\rm S}\) along which, on average, \(p  z\{x\}=2(1-p)(1-x)x\) and magnetization is conserved [Fig.~\ref{f:T}(a)]. Indeed MC simulations suggest that the triple point \(T^*\) for the full system is unique and marks the end of the active phase, similar to the triple point \(T\) in the PA.

For \((w,p)\) approaching \(T^*\) along the curve of maximum convergence times in Fig.~\ref{f:de}(b), stochastic trajectories drift towards \(x=1/2\) on the curve \(M_{\rm S}\) before slowly diffusing to consensus with increasingly small drift velocities.
At \(T^*\), the drift velocity is effectively zero, and for higher values of \(w\) the stochastic trajectories drift towards, instead of away from, the consensus states. This behavior agrees qualitatively with the description given by the PA for the thermodynamic limit, in that the whole \(M^{\rm T}_{\rm E}\) is formed by equilibria when the change of stability of the equilibrium point on \(M_{\rm E}\) occurs. 

In contrast to \(T\) in the PA description, the steady-state subensemble mean degrees of the full system are not equal at \(T^*\), see Fig.~\ref{f:T}(b). This is to be expected, because although in the approximate pairwise description zero drift on the slow manifold entails equal subensemble mean degrees, these are in general independent constraints, with the former occurring at a single point in parameter space for a given \(\langle k \rangle\). Consequently, the asymmetric VM in the full system generally cannot fully emulate the DE of the classic VM, featuring only a "weak" triple point \(T^*\) as a ghost remnant of \(T\) in the approximate pairwise description.

While the shape of \(M_{\rm S}\) at \(T^*\) encodes the dynamical aspect of a line of equilibria (the global balance of transmission and relaxation events), the associated \emph{stochastic} features should also emerge. In the reduced description 
of a one-dimesional RW along \(M_{\rm S}\), these include the fixation probability \(\pi_{\rm A}\{X_0,N\}\approx X_0/N\) for all initial \(X_0\) considered [see Eq.~(\ref{e:pi})], as well as a linear scaling of convergence times with system size. 
However, fluctuations transversal to \(M_{\rm S}\) hinder a quantitative description with frameworks like Eq.~(\ref{e:me}), as they do away from \(T^*\) in the active phase. Still, simulations reveal that indeed fixation probabilities for the approximate \(T^*\) do not display the convergence towards 0 or 1 values for increasing system sizes observed for DEs outside \(T^*\) [Fig.~\ref{f:T}(c)]. Furthermore, the observed scaling of convergence times with system size is only weakly sublinear at \(T^*\) [Fig.~\ref{f:T}(d)]. These are indications, at the level of the stochastic properties of the system, of the vicinity of a point where transmission and relaxation events balance out along \(M_{\rm S}\).

\section{Summary and outlook}\label{s:so}
The CP and the VM on adaptive networks are two paradigms of coevolutionary dynamics whose nodes cycle in binary state space while link rewiring promotes homophilic interactions. State dynamics also promote local consensus by letting a node state propagate to its neighbors. These two different mechanisms of generating concordant links, social contagion and selective interactions, are assigned to specific node states in the adaptive CP but they become 
entangled in the symmetric VM, which has full node state symmetry. 

The asymmetric VM proposed here is a modification of the adaptive CP, whose interpretation in the context of opinion dynamics breaks the symmetry of the VM by assigning to each node type a different strategy to promote consensus. B-nodes represent orthodox opinion holders that spread their opinion via social pressure and adopt a segregationist strategy, rewiring their connections in search of their peers. A-nodes represent heterodox opinion holders that relax to the orthodox opinion according to its representation in the population and adopt a proselytic strategy, converting their B-node neighbors through personal interactions. More formally, the assignment of these two ensemble specific strategies and adoption rules amounts to letting coevolutionary link-update dynamics of B-nodes coexist and compete with mean-field node-update dynamics of A-nodes.

We give a full description of the phase diagram of the asymmetric VM, using the standard pair approximation equations and assessing their performance by comparison with stochastic simulations. Not surprisingly, consensus is favored with regard to the symmetric VM, in the sense that the phase diagram is dominated by frozen phases of full consensus, including a bistable phase. We then focus on the active phase and its DE, to describe the stochastic properties of the corresponding metastable state in finite-size networks. We find that, in contrast to the symmetric VM, the final consensus state in network dynamics is, for sufficiently large system sizes, predetermined by the system's parameters and independent of initial conditions. We also find that rewiring favors consensus, with the scaling of convergence times with system size changing from exponential for  \(w=0\) to  sublinear for \(w \neq 0\).

In the PA description, the DE sits on a heteroclinic orbit connecting the two consensus states that behaves like an attracting slow manifold. Simulations reveal that this slow manifold approximates a curve that plays a similar role for the full system. 
As in the symmetric VM, the metastable regime can be qualitatively characterized by reducing system dynamics to a RW along this one-dimensional slow manifold. The latter is a line of equilibria in the symmetric case, where dynamics are trivial enough to preserve magnetization even during the system's transient. The presence of a stable DE in the asymmetric case entangles transversal with tangential fluctuations and complicates a quantitative stochastic description of metastability using this reduction as an approximation. Encouraging analytic results in that direction have been obtained for fluctuations around the DE of a slow manifold \cite{Constable2013}. However to fully characterize metastability, one would need to extend existing approaches to the entire slow manifold (see also \cite{Roberts2008}).

The PA of the asymmetric VM yields a triple point in parameter space where \(w \neq 0\) and
the system in DE emulates the link-update version of the classic VM without rewiring. This can be understood as a particular choice of parameters that restores the symmetry of the two node states.
For the full system, a similar point is identified that partially replicates the phenomenology of the triple point in the PA: While it features an attracting curve along which magnetization is conserved, the two node ensembles in DE do not display equal mean degrees. The study of these subensemble steady-state degree distributions has been left for future work, and can be addressed using the analytic framework developed in \cite{WielandEPL2012,WielandEPJ2012} together with systematic simulations. Preliminary results moreover indicate that for the symmetric VM, subensemble degree distributions remain the same as the magnetization changes along the line of equilibria. The aforementioned framework, initially designed to detect and describe pointlike DEs, could be extended to analytically support this observation.

In summary, the asymmetric VM under consideration has a straightforward interpretation in the context of opinion dynamics and it exhibits new features, both in the thermodynamic limit and in finite-size networks. A precise analytic description of its stochastic properties requires going beyond the methods developed for the symmetric VM. This effort may be justified in the scope of exploring  quantitative models of social dynamics that, with respect to the classic VM, include more realistic ingredients. Recent controlled experiments that decouple the effects of link formation and of opinion adoption on observed homophily \cite{Centola2011} open the way to a more elaborate study of homophily and social diffusion that will call upon more sophisticated models than the ones explored so far.

\begin{acknowledgments}
Financial support from the Portuguese
Foundation for Science and Technology (FCT) under
Contract POCTI/ISFL/2/261 is gratefully acknowledged.
SW was also supported by FCT under Grant No. SFRH/BD/45179/2008.
\end{acknowledgments}
\appendix
\section{A-consensus as a fixed point}\label{s:app}
Expressing Eqs.~(\ref{e:pa}) in spherical coordinates \(\theta, \phi \in [0,\pi/2]\) and \(r \in [0,1]\) with \(x=1-r\sin(\theta)\cos(\phi)\), \(y=\langle k \rangle/2- r\sin(\theta)\sin(\phi)\), and \(z=r\cos(\theta)\), one is interested in the flow's behavior in the vicinity of A-consensus, i.e., for \(r\rightarrow 0\). 

In these coordinates, the singularity at A-consensus can be regularized, and we find that  \(\dot{r}=0\) for \(r=0\) and all angular coordinates. Hence the sphere octant \(\Sigma\) at \(r=0\) is invariant and represents A-consensus, the latter of which is then indeed a steady state. Parameter regions of stable A-consensus can consequently be determined by identifying stable fixed points \((\theta^*,\phi^*)\) of the angular flow on \(\Sigma\), in conjunction with demanding stability of the radial flow at \((\theta^*,\phi^*)\). In the following, "fixed point" will be exclusively used in the context of the two-dimensional angular flow on \(\Sigma\), whereas A-consensus in the full system is referred to as such. 

It is straightforward to classify the parameter dependence of the angular flow  into six typical phase portraits, each of which yielding three fixed points at the boundary \(\phi=\pi/2\) and none, one or two in the interior of \(\Sigma\). With these phase portraits at hand, the stability of fixed points is separately investigated for those two regions.

At the boundary \(\phi=\pi/2\), the angular flow has fixed points \((\theta^*,\phi^*)\) at \((\arccos\{2/\sqrt{5}\},\pi/2)\), \((\pi/4,\pi/2)\), and \((\pi/2,\pi/2)\). The first one is unstable, whereas the remaining two are non-hyperbolic with one zero and one negative eigenvalue each. The radial flow is unstable at \((\pi/2,\pi/2)\) and stable at \((\pi/4,\pi/2)\), so that only the latter fixed point is of further interest. Aforementioned phase portraits reveal it to be stable as soon as there are two [for \(p<2/(1+\langle k \rangle)\)] or no [for \(p>2/(1+\langle k \rangle)\)] fixed points in the interior of \(\Sigma\). For both cases, this criterion translates into a comparison of tangent slopes of respective nullclines at \((\pi/4,\pi/2)\) and yields
\[p>(2-3w)/(1-w)\]
as a sufficient condition for stable A-consensus. 

Consequently for \(p<(2-3w)/(1-w)\), there is exactly one fixed point in the interior of \(\Sigma\), and it follows from aforementioned phase portraits that it is always stable. It can moreover be shown that the radial flow at this fixed point is stable if 
\[p>2/(1+\langle k \rangle) \ ,\]
so that overall, A-consensus is stable for 
\[p>\min\{2/(1+\langle k \rangle),(2-3w)/(1-w)\} \ .\]
Note that there can be coexisting stable fixed points for the angular flow - one at \((\pi/4,\pi/2)\) (automatically implying stable A-consensus) and one in the interior of \(\Sigma\). For the \(\langle k \rangle\) chosen here, the basin of attraction of the former is significantly smaller than that of the latter. If the radial flow is unstable at the interior fixed point, trajectories in the vicinity of A-consensus usually get attracted to those angular coordinates and are consequently repelled from A-consensus in the \(r\)-direction. This is the case for low \(p\) and large \(w\), so that reaching A-consensus for that parameter region requires careful selection of initial conditions [Fig.~\ref{f:mG}(b)]. If however the radial flow is also stable at the interior stable fixed point, then there are two routes towards A-consensus, characterized by the two different angles under which the trajectory approaches it. 
\section{Splitting probabilities}\label{s:app2}
In what follows, a condensed version of the relevant section in \cite{vanKampen2007} is laid out: Considering a general one-step process in the positive integer variable \(0\leq i \leq N\), transition rates \(g_{i}\) (mediating a gain in \(i\)), \(r_{i}\) (descending \(i\)) and the two absorbing integer boundaries \(0\) and \(N\geq 0\) are defined. Denoting by \(\pi_{i}\) the splitting probability for the random walker to reach \(N\) before \(0\) when starting from \(i\), 
\begin{equation*}
\pi_{i}=\frac{g_i}{g_i+r_i}\pi_{i+1}+\frac{r_i}{g_i+r_i}\pi_{i-1} 
\end{equation*}
and equivalently
\begin{equation}\label{app:splitt}
0=g_i(\pi_{i+1}-\pi_{i})+r_i(\pi_{i-1}-\pi_{i})
\end{equation}
obviously hold for \(2\leq i \leq(N-2)\) as a recursive definition of \(\pi_{i}\). As the boundary conditions are \(\pi_{0}=0\) and \(\pi_{N}=1\), Eq.~(\ref{app:splitt}) extends its validity to \(1\leq i \leq (N-1)\). For the same interval of \(i\), setting \(\Delta_i\equiv(\pi_{i+1}-\pi_{i})\)
yields \(g_i\Delta_i=r_i\Delta_{i-1}\), so that 
\[\Delta_i=\prod_{j=1}^i \frac{r_j}{g_j}\Delta_0\]
with \(\Delta_0=\pi_{1}\). It follows that
\begin{equation*}
\pi_{i}=\sum\limits_{j=0}^{i-1}\Delta_j=\pi_{1}+\sum\limits_{j=1}^{i-1}\prod_{k=1}^{j} \frac{r_k}{g_k}\pi_{1} \, .
\end{equation*}
Considering that \(\pi_{N}=1\) delivers \(\pi_{1}\), so that finally
\begin{align}\label{e:APPfinalsplit}
\pi_{i} &=\frac{1+\sum_{j=1}^{i-1}\prod_{k=1}^{j} r_k/g_k}{1+\sum_{j=1}^{N-1}\prod_{k=1}^{j} r_k/g_k}\, 
\end{align}
Inserting the transition rates of Eq.~(\ref{e:me})
readily yields the splitting probabilities for the asymmetric VM.
\section{Mean first-passage times}\label{s:app3}
In the same recursive manner as in Sec.~\ref{s:app2}, one can compute the mean first-passage time \(\tau_i\) of the random walker to hit coordinate \(N\) before \(0\) when starting at coordinate \(0\leq i\leq N\). Within the first time step \(\Delta t\), the random walker jumps to \((i+1)\) with probability \( g_i\Delta t\), to \((i-1)\) with probability \( r_i\Delta t\), and stays at \(i\) with probability \((1- g_i\Delta t- r_i\Delta t)\). Then \(\pi_{i}(\tau_i-\Delta t)\) is equal to a weighted sum of \(\pi_{i+1}\tau_{i+1}\), \(\pi_{i}\tau_{i}\) and \(\pi_{i-1}\tau_{i-1}\), with the weights being aforementioned transition probabilities to the respective coordinates. Introducing \(\nu_i\equiv \pi_{i} \tau_i\), one obtains
\begin{equation*}
\nu_i-\pi_{i}\Delta t= \nu_{i+1}g_i\Delta t+\nu_{i-1}r_i\Delta t+\nu_i(1- g_i\Delta t- r_i\Delta t)
\end{equation*}
and, through redefining \(\Delta_i\equiv \nu_{i+1}-\nu_i\), 
\begin{equation*}
\Delta_i=-\frac{\pi_i}{g_i}+\frac{r_i}{g_i}\Delta_{i-1}\, .
\end{equation*}
It is straightforward to check that the latter recursive equation can be given the closed form \cite{vanKampen2007}
\begin{equation*}
\Delta_i=\nu_1\prod_{k=1}^{i}\frac{r_k}{g_k}-\sum\limits_{j=1}^{i}\frac{\pi_j}{r_j}\prod_{k=j}^{i}\frac{r_k}{g_k}\, ,
\end{equation*}
so that then
\begin{align}\label{e:APPpitau2}
\nu_i=\sum\limits_{j=0}^{i-1}\Delta_j=\nu_1\left(1+\sum\limits_{j=1}^{i-1}\prod\limits_{k=1}^{j}\frac{r_k}{g_k}\right)-\sum\limits_{j=1}^{i-1}\sum\limits_{k=1}^{j}\frac{\pi_k}{r_k}\prod_{l=k}^{j}\frac{r_l}{g_l}\, .
\end{align}
Since \(\nu_N=0\) due to \(\tau_N=0\), \(\nu_1\) can be calculated from Eq.~(\ref{e:APPpitau2}) as
\begin{equation}\label{e:APPpitau3}
\nu_1=\frac{\sum_{j=1}^{N-1}\sum_{k=1}^{j}\pi_k/r_k\prod_{l=k}^{j}r_l/g_l}{1+\sum_{j=1}^{N-1}\prod_{k=1}^{j}r_k/g_k} \, .
\end{equation}
Inserting Eq.~(\ref{e:APPpitau3}) into Eq.~(\ref{e:APPpitau2}) while considering Eq.~(\ref{e:APPfinalsplit}) finally yields
\begin{align}\label{e:APPpitau4}
\nu_i&=\pi_i \sum\limits_{j=1}^{N-1}\sum\limits_{k=1}^{j}\frac{\pi_k}{r_k}\prod\limits_{l=k}^{j}\frac{r_l}{g_l}-\sum\limits_{j=1}^{i-1}\sum\limits_{k=1}^{j}\frac{\pi_k}{r_k}\prod_{l=k}^{j}\frac{r_l}{g_l}\nonumber\\
&=\pi_i \sum\limits_{j=i}^{N-1}\sum\limits_{k=1}^{j}\frac{\pi_k}{r_k}\prod\limits_{l=k}^{j}\frac{r_l}{g_l}-(1-\pi_i)\sum\limits_{j=1}^{i-1}\sum\limits_{k=1}^{j}\frac{\pi_k}{r_k}\prod_{l=k}^{j}\frac{r_l}{g_l} \,  ,
\end{align}
where for large \(N\) and \(i\), the second equality is to be preferred to ensure fast computation of \(\nu_i\). 

To calculate the mean first-passage time \(\tau_i^{'}\) for hitting coordinate \(0\) first when starting from \(i\), the variable \(\nu_i^{'}\equiv(1-\pi_i)\tau_i^{'}\) is introduced. Then \(\nu_i^{'}\) is easily obtained from Eq.~(\ref{e:APPpitau4}) by reversing the direction of the considered first passage, so that
\begin{align}\label{e:APPpitau5}
\nu_i^{'}=&(1-\pi_i )\sum\limits_{j=N-i}^{N-1}\sum\limits_{k=1}^{j}\frac{\pi_{N-k}}{g_{N-k}}\prod\limits_{l=k}^{j}\frac{g_{N-l}}{r_{N-l}}\nonumber \\
          &-\pi_i\sum\limits_{j=1}^{N-i-1}\sum\limits_{k=1}^{j}\frac{\pi_{N-k}}{g_{N-k}}\prod_{l=k}^{j}\frac{g_{N-l}}{r_{N-l}} \, .
\end{align}
Finally,
\begin{equation}\label{e:APPpitau6}
\tau_i=\nu_i+\nu_i^{'}
\end{equation}
is the mean passage time for hitting either of the interval boundaries when starting from coordinate \(i\).

\bibliography{PBSpre2013}

\begin{thebibliography}{30}%
\makeatletter
\providecommand \@ifxundefined [1]{%
 \@ifx{#1\undefined}
}%
\providecommand \@ifnum [1]{%
 \ifnum #1\expandafter \@firstoftwo
 \else \expandafter \@secondoftwo
 \fi
}%
\providecommand \@ifx [1]{%
 \ifx #1\expandafter \@firstoftwo
 \else \expandafter \@secondoftwo
 \fi
}%
\providecommand \natexlab [1]{#1}%
\providecommand \enquote  [1]{``#1''}%
\providecommand \bibnamefont  [1]{#1}%
\providecommand \bibfnamefont [1]{#1}%
\providecommand \citenamefont [1]{#1}%
\providecommand \href@noop [0]{\@secondoftwo}%
\providecommand \href [0]{\begingroup \@sanitize@url \@href}%
\providecommand \@href[1]{\@@startlink{#1}\@@href}%
\providecommand \@@href[1]{\endgroup#1\@@endlink}%
\providecommand \@sanitize@url [0]{\catcode `\\12\catcode `\$12\catcode
  `\&12\catcode `\#12\catcode `\^12\catcode `\_12\catcode `\%12\relax}%
\providecommand \@@startlink[1]{}%
\providecommand \@@endlink[0]{}%
\providecommand \url  [0]{\begingroup\@sanitize@url \@url }%
\providecommand \@url [1]{\endgroup\@href {#1}{\urlprefix }}%
\providecommand \urlprefix  [0]{URL }%
\providecommand \Eprint [0]{\href }%
\providecommand \doibase [0]{http://dx.doi.org/}%
\providecommand \selectlanguage [0]{\@gobble}%
\providecommand \bibinfo  [0]{\@secondoftwo}%
\providecommand \bibfield  [0]{\@secondoftwo}%
\providecommand \translation [1]{[#1]}%
\providecommand \BibitemOpen [0]{}%
\providecommand \bibitemStop [0]{}%
\providecommand \bibitemNoStop [0]{.\EOS\space}%
\providecommand \EOS [0]{\spacefactor3000\relax}%
\providecommand \BibitemShut  [1]{\csname bibitem#1\endcsname}%
\let\auto@bib@innerbib\@empty
\bibitem [{\citenamefont {Dorogovtsev}\ and\ \citenamefont
  {Mendes}(2003)}]{Dorogovtsev2003}%
  \BibitemOpen
  \bibfield  {author} {\bibinfo {author} {\bibfnamefont {S.~N.}\ \bibnamefont
  {Dorogovtsev}}\ and\ \bibinfo {author} {\bibfnamefont {J.~F.~F.}\
  \bibnamefont {Mendes}},\ }\href@noop {} {\emph {\bibinfo {title} {Evolution
  of Networks: From Biological Nets to the Internet and WWW}}}\ (\bibinfo
  {publisher} {Oxford University Press},\ \bibinfo {address} {New York, NY,
  USA},\ \bibinfo {year} {2003})\BibitemShut {NoStop}%
\bibitem [{\citenamefont {Newman}(2003)}]{NewmanSIAM2003}%
  \BibitemOpen
  \bibfield  {author} {\bibinfo {author} {\bibfnamefont {M.~E.~J.}\
  \bibnamefont {Newman}},\ }\href {\doibase 10.1137/S003614450342480}
  {\bibfield  {journal} {\bibinfo  {journal} {SIAM Review}\ }\textbf {\bibinfo
  {volume} {45}},\ \bibinfo {pages} {167} (\bibinfo {year} {2003})}\BibitemShut
  {NoStop}%
\bibitem [{\citenamefont {Gross}\ and\ \citenamefont
  {Blasius}(2008)}]{Blasius2008}%
  \BibitemOpen
  \bibfield  {author} {\bibinfo {author} {\bibfnamefont {T.}~\bibnamefont
  {Gross}}\ and\ \bibinfo {author} {\bibfnamefont {B.}~\bibnamefont
  {Blasius}},\ }\href@noop {} {\bibfield  {journal} {\bibinfo  {journal} {J. R.
  Soc. Interface}\ }\textbf {\bibinfo {volume} {5}},\ \bibinfo {pages} {259}
  (\bibinfo {year} {2008})}\BibitemShut {NoStop}%
\bibitem [{\citenamefont {Gross}\ and\ \citenamefont
  {Sayama}(2009)}]{Gross2009}%
  \BibitemOpen
  \bibinfo {editor} {\bibfnamefont {T.}~\bibnamefont {Gross}}\ and\ \bibinfo
  {editor} {\bibfnamefont {H.}~\bibnamefont {Sayama}},\ eds.,\ \href@noop {}
  {\emph {\bibinfo {title} {{Adaptive networks: Theory, Models and
  Applications}}}}\ (\bibinfo  {publisher} {Springer},\ \bibinfo {address} {New
  York},\ \bibinfo {year} {2009})\BibitemShut {NoStop}%
\bibitem [{\citenamefont {Sayama}\ \emph {et~al.}(2013)\citenamefont {Sayama},
  \citenamefont {Pestov}, \citenamefont {Schmidt}, \citenamefont {Bush},
  \citenamefont {Wong}, \citenamefont {Yamanoi},\ and\ \citenamefont
  {Gross}}]{Sayama2013}%
  \BibitemOpen
  \bibfield  {author} {\bibinfo {author} {\bibfnamefont {H.}~\bibnamefont
  {Sayama}}, \bibinfo {author} {\bibfnamefont {P.}~\bibnamefont {Pestov}},
  \bibinfo {author} {\bibfnamefont {J.}~\bibnamefont {Schmidt}}, \bibinfo
  {author} {\bibfnamefont {B.~J.}\ \bibnamefont {Bush}}, \bibinfo {author}
  {\bibfnamefont {C.}~\bibnamefont {Wong}}, \bibinfo {author} {\bibfnamefont
  {J.}~\bibnamefont {Yamanoi}}, \ and\ \bibinfo {author} {\bibfnamefont
  {T.}~\bibnamefont {Gross}},\ }\href {\doibase 10.1016/j.camwa.2012.12.005}
  {\bibfield  {journal} {\bibinfo  {journal} {Computers and Mathematics with
  Applications}\ ,\ } (\bibinfo {year} {2013})}\BibitemShut {NoStop}%
\bibitem [{\citenamefont {Vazquez}\ and\ \citenamefont
  {Egu\'iluz}(2008)}]{VazquezNJP2008}%
  \BibitemOpen
  \bibfield  {author} {\bibinfo {author} {\bibfnamefont {F.}~\bibnamefont
  {Vazquez}}\ and\ \bibinfo {author} {\bibfnamefont {V.~M.}\ \bibnamefont
  {Egu\'iluz}},\ }\href@noop {} {\bibfield  {journal} {\bibinfo  {journal} {New
  Journal of Physics}\ }\textbf {\bibinfo {volume} {10}},\ \bibinfo {pages}
  {063011} (\bibinfo {year} {2008})}\BibitemShut {NoStop}%
\bibitem [{\citenamefont {Gross}\ \emph {et~al.}(2006)\citenamefont {Gross},
  \citenamefont {Dommar},\ and\ \citenamefont {Blasius}}]{GrossPRL2006}%
  \BibitemOpen
  \bibfield  {author} {\bibinfo {author} {\bibfnamefont {T.}~\bibnamefont
  {Gross}}, \bibinfo {author} {\bibfnamefont {C.~J.}\ \bibnamefont {Dommar}}, \
  and\ \bibinfo {author} {\bibfnamefont {B.}~\bibnamefont {Blasius}},\ }\href
  {\doibase 10.1103/PhysRevLett.96.208701} {\bibfield  {journal} {\bibinfo
  {journal} {Phys. Rev. Lett.}\ }\textbf {\bibinfo {volume} {96}},\ \bibinfo
  {pages} {208701} (\bibinfo {year} {2006})}\BibitemShut {NoStop}%
\bibitem [{\citenamefont {Kimura}\ and\ \citenamefont
  {Hayakawa}(2008)}]{KimuraPRE2008}%
  \BibitemOpen
  \bibfield  {author} {\bibinfo {author} {\bibfnamefont {D.}~\bibnamefont
  {Kimura}}\ and\ \bibinfo {author} {\bibfnamefont {Y.}~\bibnamefont
  {Hayakawa}},\ }\href {\doibase 10.1103/PhysRevE.78.016103} {\bibfield
  {journal} {\bibinfo  {journal} {Phys. Rev. E}\ }\textbf {\bibinfo {volume}
  {78}},\ \bibinfo {pages} {016103} (\bibinfo {year} {2008})}\BibitemShut
  {NoStop}%
\bibitem [{\citenamefont {B\"ohme}\ and\ \citenamefont
  {Gross}(2011)}]{BoehmePRE2011}%
  \BibitemOpen
  \bibfield  {author} {\bibinfo {author} {\bibfnamefont {G.~A.}\ \bibnamefont
  {B\"ohme}}\ and\ \bibinfo {author} {\bibfnamefont {T.}~\bibnamefont
  {Gross}},\ }\href {\doibase 10.1103/PhysRevE.83.035101} {\bibfield  {journal}
  {\bibinfo  {journal} {Phys. Rev. E}\ }\textbf {\bibinfo {volume} {83}},\
  \bibinfo {pages} {035101} (\bibinfo {year} {2011})}\BibitemShut {NoStop}%
\bibitem [{\citenamefont {Vazquez}\ \emph {et~al.}(2008)\citenamefont
  {Vazquez}, \citenamefont {Egu\'iluz},\ and\ \citenamefont
  {Miguel}}]{VazquezPRL2008}%
  \BibitemOpen
  \bibfield  {author} {\bibinfo {author} {\bibfnamefont {F.}~\bibnamefont
  {Vazquez}}, \bibinfo {author} {\bibfnamefont {V.~M.}\ \bibnamefont
  {Egu\'iluz}}, \ and\ \bibinfo {author} {\bibfnamefont {M.~S.}\ \bibnamefont
  {Miguel}},\ }\href {\doibase 10.1103/PhysRevLett.100.108702} {\bibfield
  {journal} {\bibinfo  {journal} {Phys. Rev. Lett.}\ }\textbf {\bibinfo
  {volume} {100}},\ \bibinfo {pages} {108702} (\bibinfo {year}
  {2008})}\BibitemShut {NoStop}%
\bibitem [{\citenamefont {Sood}\ and\ \citenamefont
  {Redner}(2005)}]{SoodPRL2005}%
  \BibitemOpen
  \bibfield  {author} {\bibinfo {author} {\bibfnamefont {V.}~\bibnamefont
  {Sood}}\ and\ \bibinfo {author} {\bibfnamefont {S.}~\bibnamefont {Redner}},\
  }\href {\doibase 10.1103/PhysRevLett.94.178701} {\bibfield  {journal}
  {\bibinfo  {journal} {Phys. Rev. Lett.}\ }\textbf {\bibinfo {volume} {94}},\
  \bibinfo {pages} {178701} (\bibinfo {year} {2005})}\BibitemShut {NoStop}%
\bibitem [{\citenamefont {Nardini}\ \emph {et~al.}(2008)\citenamefont
  {Nardini}, \citenamefont {Kozma},\ and\ \citenamefont
  {Barrat}}]{Nardini2008}%
  \BibitemOpen
  \bibfield  {author} {\bibinfo {author} {\bibfnamefont {C.}~\bibnamefont
  {Nardini}}, \bibinfo {author} {\bibfnamefont {B.}~\bibnamefont {Kozma}}, \
  and\ \bibinfo {author} {\bibfnamefont {A.}~\bibnamefont {Barrat}},\ }\href
  {\doibase 10.1103/PhysRevLett.100.158701} {\bibfield  {journal} {\bibinfo
  {journal} {Phys. Rev. Lett.}\ }\textbf {\bibinfo {volume} {100}},\ \bibinfo
  {pages} {158701} (\bibinfo {year} {2008})}\BibitemShut {NoStop}%
\bibitem [{\citenamefont {Holley}\ and\ \citenamefont
  {Liggett}(1975)}]{Holley1975}%
  \BibitemOpen
  \bibfield  {author} {\bibinfo {author} {\bibfnamefont {R.~A.}\ \bibnamefont
  {Holley}}\ and\ \bibinfo {author} {\bibfnamefont {T.~M.}\ \bibnamefont
  {Liggett}},\ }\href@noop {} {\bibfield  {journal} {\bibinfo  {journal} {Ann.
  Probab}\ }\textbf {\bibinfo {volume} {3}},\ \bibinfo {pages} {643} (\bibinfo
  {year} {1975})}\BibitemShut {NoStop}%
\bibitem [{\citenamefont {Demirel}\ \emph {et~al.}(2013)\citenamefont
  {Demirel}, \citenamefont {Vazquez}, \citenamefont {B\"ohme},\ and\
  \citenamefont {Gross}}]{Demirel2013}%
  \BibitemOpen
  \bibfield  {author} {\bibinfo {author} {\bibfnamefont {G.}~\bibnamefont
  {Demirel}}, \bibinfo {author} {\bibfnamefont {F.}~\bibnamefont {Vazquez}},
  \bibinfo {author} {\bibfnamefont {G.~A.}\ \bibnamefont {B\"ohme}}, \ and\
  \bibinfo {author} {\bibfnamefont {T.}~\bibnamefont {Gross}},\ }\href@noop {}
  {\bibfield  {journal} {\bibinfo  {journal} {arXiv:1211.0449 [nlin.AO]}\ }
  (\bibinfo {year} {2013})}\BibitemShut {NoStop}%
\bibitem [{\citenamefont {Suchecki}\ \emph {et~al.}(2005)\citenamefont
  {Suchecki}, \citenamefont {Egu\'iluz},\ and\ \citenamefont
  {Miguel}}]{SucheckiEPL2004}%
  \BibitemOpen
  \bibfield  {author} {\bibinfo {author} {\bibfnamefont {K.}~\bibnamefont
  {Suchecki}}, \bibinfo {author} {\bibfnamefont {V.~M.}\ \bibnamefont
  {Egu\'iluz}}, \ and\ \bibinfo {author} {\bibfnamefont {M.~S.}\ \bibnamefont
  {Miguel}},\ }\href@noop {} {\bibfield  {journal} {\bibinfo  {journal} {EPL}\
  }\textbf {\bibinfo {volume} {69}},\ \bibinfo {pages} {228} (\bibinfo {year}
  {2005})}\BibitemShut {NoStop}%
\bibitem [{\citenamefont {Galam}(2005)}]{Galam2005}%
  \BibitemOpen
  \bibfield  {author} {\bibinfo {author} {\bibfnamefont {S.}~\bibnamefont
  {Galam}},\ }\href@noop {} {\bibfield  {journal} {\bibinfo  {journal} {EPL}\
  }\textbf {\bibinfo {volume} {70}},\ \bibinfo {pages} {705} (\bibinfo {year}
  {2005})}\BibitemShut {NoStop}%
\bibitem [{\citenamefont {Galam}(2008)}]{Galam2008}%
  \BibitemOpen
  \bibfield  {author} {\bibinfo {author} {\bibfnamefont {S.}~\bibnamefont
  {Galam}},\ }\href@noop {} {\bibfield  {journal} {\bibinfo  {journal}
  {International Journal of Modern Physics C}\ }\textbf {\bibinfo {volume}
  {19}},\ \bibinfo {pages} {409} (\bibinfo {year} {2008})}\BibitemShut
  {NoStop}%
\bibitem [{\citenamefont {Williams}\ and\ \citenamefont
  {Bjerknes}(1972)}]{Williams1972}%
  \BibitemOpen
  \bibfield  {author} {\bibinfo {author} {\bibfnamefont {T.}~\bibnamefont
  {Williams}}\ and\ \bibinfo {author} {\bibfnamefont {R.}~\bibnamefont
  {Bjerknes}},\ }\href {\doibase 10.1038/236019a0} {\bibfield  {journal}
  {\bibinfo  {journal} {Nature}\ }\textbf {\bibinfo {volume} {236}},\ \bibinfo
  {pages} {19} (\bibinfo {year} {1972})}\BibitemShut {NoStop}%
\bibitem [{\citenamefont {S\'anchez}\ \emph {et~al.}(2002)\citenamefont
  {S\'anchez}, \citenamefont {L\'opez},\ and\ \citenamefont
  {Rodr\'iguez}}]{SanchezPRL2002}%
  \BibitemOpen
  \bibfield  {author} {\bibinfo {author} {\bibfnamefont {A.~D.}\ \bibnamefont
  {S\'anchez}}, \bibinfo {author} {\bibfnamefont {J.~M.}\ \bibnamefont
  {L\'opez}}, \ and\ \bibinfo {author} {\bibfnamefont {M.~A.}\ \bibnamefont
  {Rodr\'iguez}},\ }\href {\doibase 10.1103/PhysRevLett.88.048701} {\bibfield
  {journal} {\bibinfo  {journal} {Phys. Rev. Lett.}\ }\textbf {\bibinfo
  {volume} {88}},\ \bibinfo {pages} {048701} (\bibinfo {year}
  {2002})}\BibitemShut {NoStop}%
\bibitem [{\citenamefont {Zschaler}\ \emph {et~al.}(2012)\citenamefont
  {Zschaler}, \citenamefont {B\"ohme}, \citenamefont {Sei\ss{}inger},
  \citenamefont {Huepe},\ and\ \citenamefont {Gross}}]{ZschalerPRE2012}%
  \BibitemOpen
  \bibfield  {author} {\bibinfo {author} {\bibfnamefont {G.}~\bibnamefont
  {Zschaler}}, \bibinfo {author} {\bibfnamefont {G.~A.}\ \bibnamefont
  {B\"ohme}}, \bibinfo {author} {\bibfnamefont {M.}~\bibnamefont
  {Sei\ss{}inger}}, \bibinfo {author} {\bibfnamefont {C.}~\bibnamefont
  {Huepe}}, \ and\ \bibinfo {author} {\bibfnamefont {T.}~\bibnamefont
  {Gross}},\ }\href {\doibase 10.1103/PhysRevE.85.046107} {\bibfield  {journal}
  {\bibinfo  {journal} {Phys. Rev. E}\ }\textbf {\bibinfo {volume} {85}},\
  \bibinfo {pages} {046107} (\bibinfo {year} {2012})}\BibitemShut {NoStop}%
\bibitem [{\citenamefont {Wieland}\ \emph
  {et~al.}(2012{\natexlab{a}})\citenamefont {Wieland}, \citenamefont {Parisi},\
  and\ \citenamefont {Nunes}}]{WielandEPJ2012}%
  \BibitemOpen
  \bibfield  {author} {\bibinfo {author} {\bibfnamefont {S.}~\bibnamefont
  {Wieland}}, \bibinfo {author} {\bibfnamefont {A.}~\bibnamefont {Parisi}}, \
  and\ \bibinfo {author} {\bibfnamefont {A.}~\bibnamefont {Nunes}},\ }\href
  {\doibase 10.1140/epjst/e2012-01656-5} {\bibfield  {journal} {\bibinfo
  {journal} {EPJ-ST}\ }\textbf {\bibinfo {volume} {212}},\ \bibinfo {pages}
  {99} (\bibinfo {year} {2012}{\natexlab{a}})}\BibitemShut {NoStop}%
\bibitem [{\citenamefont {Gillespie}(1976)}]{Gillespie1976}%
  \BibitemOpen
  \bibfield  {author} {\bibinfo {author} {\bibfnamefont {D.~T.}\ \bibnamefont
  {Gillespie}},\ }\href {\doibase 10.1016/0021-9991(76)90041-3} {\bibfield
  {journal} {\bibinfo  {journal} {J. Comput. Phys.}\ }\textbf {\bibinfo
  {volume} {22}},\ \bibinfo {pages} {403} (\bibinfo {year} {1976})}\BibitemShut
  {NoStop}%
\bibitem [{\citenamefont {Erd\H{o}s}\ and\ \citenamefont
  {R\'{e}nyi}(1960)}]{Erdos1960}%
  \BibitemOpen
  \bibfield  {author} {\bibinfo {author} {\bibfnamefont {P.}~\bibnamefont
  {Erd\H{o}s}}\ and\ \bibinfo {author} {\bibfnamefont {A.}~\bibnamefont
  {R\'{e}nyi}},\ }in\ \href@noop {} {\emph {\bibinfo {booktitle} {Publication
  of the Mathematical Institute of the Hungarian Academy of Sciences}}}\
  (\bibinfo {year} {1960})\ pp.\ \bibinfo {pages} {17--61}\BibitemShut
  {NoStop}%
\bibitem [{\citenamefont {van Kampen}(2007)}]{vanKampen2007}%
  \BibitemOpen
  \bibfield  {author} {\bibinfo {author} {\bibfnamefont {N.}~\bibnamefont {van
  Kampen}},\ }\href@noop {} {\emph {\bibinfo {title} {Stochastic Processes in
  Physics and Chemistry, 3rd Edition}}}\ (\bibinfo  {publisher} {North
  Holland},\ \bibinfo {year} {2007})\BibitemShut {NoStop}%
\bibitem [{\citenamefont {Rogers}\ and\ \citenamefont
  {Gross}(2013)}]{Rogers2013}%
  \BibitemOpen
  \bibfield  {author} {\bibinfo {author} {\bibfnamefont {T.}~\bibnamefont
  {Rogers}}\ and\ \bibinfo {author} {\bibfnamefont {T.}~\bibnamefont {Gross}},\
  }\href@noop {} {\bibfield  {journal} {\bibinfo  {journal} {arXiv:1304.4742
  [physics.soc-ph]}\ } (\bibinfo {year} {2013})}\BibitemShut {NoStop}%
\bibitem [{\citenamefont {Antal}\ \emph {et~al.}(2006)\citenamefont {Antal},
  \citenamefont {Redner},\ and\ \citenamefont {Sood}}]{AntalPRL2006}%
  \BibitemOpen
  \bibfield  {author} {\bibinfo {author} {\bibfnamefont {T.}~\bibnamefont
  {Antal}}, \bibinfo {author} {\bibfnamefont {S.}~\bibnamefont {Redner}}, \
  and\ \bibinfo {author} {\bibfnamefont {V.}~\bibnamefont {Sood}},\ }\href
  {\doibase 10.1103/PhysRevLett.96.188104} {\bibfield  {journal} {\bibinfo
  {journal} {Phys. Rev. Lett.}\ }\textbf {\bibinfo {volume} {96}},\ \bibinfo
  {pages} {188104} (\bibinfo {year} {2006})}\BibitemShut {NoStop}%
\bibitem [{\citenamefont {Constable}\ \emph {et~al.}(2013)\citenamefont
  {Constable}, \citenamefont {McKane},\ and\ \citenamefont
  {Rogers}}]{Constable2013}%
  \BibitemOpen
  \bibfield  {author} {\bibinfo {author} {\bibfnamefont {G.~W.~A.}\
  \bibnamefont {Constable}}, \bibinfo {author} {\bibfnamefont {A.~J.}\
  \bibnamefont {McKane}}, \ and\ \bibinfo {author} {\bibfnamefont
  {T.}~\bibnamefont {Rogers}},\ }\href@noop {} {\bibfield  {journal} {\bibinfo
  {journal} {J. Phys. A: Math. Theor.}\ }\textbf {\bibinfo {volume} {46}},\
  \bibinfo {pages} {295002} (\bibinfo {year} {2013})}\BibitemShut {NoStop}%
\bibitem [{\citenamefont {Roberts}(2008)}]{Roberts2008}%
  \BibitemOpen
  \bibfield  {author} {\bibinfo {author} {\bibfnamefont {A.}~\bibnamefont
  {Roberts}},\ }\href {\doibase 10.1016/j.physa.2007.08.023} {\bibfield
  {journal} {\bibinfo  {journal} {Physica A}\ }\textbf {\bibinfo {volume}
  {387}},\ \bibinfo {pages} {12 } (\bibinfo {year} {2008})}\BibitemShut
  {NoStop}%
\bibitem [{\citenamefont {Wieland}\ \emph
  {et~al.}(2012{\natexlab{b}})\citenamefont {Wieland}, \citenamefont {Aquino},\
  and\ \citenamefont {Nunes}}]{WielandEPL2012}%
  \BibitemOpen
  \bibfield  {author} {\bibinfo {author} {\bibfnamefont {S.}~\bibnamefont
  {Wieland}}, \bibinfo {author} {\bibfnamefont {T.}~\bibnamefont {Aquino}}, \
  and\ \bibinfo {author} {\bibfnamefont {A.}~\bibnamefont {Nunes}},\
  }\href@noop {} {\bibfield  {journal} {\bibinfo  {journal} {EPL}\ }\textbf
  {\bibinfo {volume} {97}},\ \bibinfo {pages} {18003} (\bibinfo {year}
  {2012}{\natexlab{b}})}\BibitemShut {NoStop}%
\bibitem [{\citenamefont {Centola}(2011)}]{Centola2011}%
  \BibitemOpen
  \bibfield  {author} {\bibinfo {author} {\bibfnamefont {D.}~\bibnamefont
  {Centola}},\ }\href {\doibase 10.1126/science.1207055} {\bibfield  {journal}
  {\bibinfo  {journal} {Science}\ }\textbf {\bibinfo {volume} {334}},\ \bibinfo
  {pages} {1269} (\bibinfo {year} {2011})}\BibitemShut {NoStop}%
\end{thebibliography}%
\end{document}